\documentclass[3p,times]{elsarticle}

\usepackage{ecrc}
\usepackage{epsfig}
\usepackage{graphics}
\volume{00}

\firstpage{1}

\journalname{Nuclear Physics B - Proceedings Supplements}

\runauth{W. Rodejohann}

\jid{NUPHBP}

\jnltitlelogo{Nuclear Physics B - Proceedings Supplements}

\usepackage{amssymb}

\def\be{\begin{equation}}
\def\ee{\end{equation}}
\def\gs{\mathrel{
   \rlap{\raise 0.511ex \hbox{$>$}}{\lower 0.511ex \hbox{$\sim$}}}}
\def\ls{\mathrel{
   \rlap{\raise 0.511ex \hbox{$<$}}{\lower 0.511ex \hbox{$\sim$}}}}
\newcommand{\obb}{0\mbox{$\nu\beta\beta$}}

\newcommand{\ba}{\begin{array}{c}}
\newcommand{\baz}{\begin{array}{cc}}
\newcommand{\bad}{\begin{array}{ccc}}
\newcommand{\bav}{\begin{array}{cccc}}
\newcommand{\bea}{\begin{equation} \begin{array}{c}}
\newcommand{\eea}{ \end{array} \end{equation}}
\newcommand{\ea}{\end{array}}

\newcommand{\dma}{\mbox{$\Delta m^2_{\rm A}$}}
\newcommand{\meff}{\mbox{$\langle m_{ee} \rangle$}}

\begin{document}

\begin{frontmatter}

\dochead{}

\title{Neutrinoless Double Beta Decay in Particle Physics}

\author{Werner Rodejohann}

\address{Max--Planck--Institut f\"ur Kernphysik, 
Postfach 103980, D--69029 Heidelberg, Germany}

\begin{abstract}
 Neutrinoless double beta decay is a process of fundamental importance
for particle physics. It can be mediated by light massive Majorana
neutrinos ({\it standard interpretation}) or by something else
({\it non-standard interpretations}). We review its
dependence on the neutrino parameters, its complementarity to other
observables sensitive to neutrino mass, and emphasize its ability to
distinguish different neutrino mass models. Then we discuss  
mechanisms different from light Majorana neutrino exchange, and show 
what can be learned from those and how they could be tested.
\end{abstract}


\end{frontmatter}

\section{\label{sec:intro}Introduction}
Neutrinoless double beta decay ($\obb$) experiments \cite{exp} are 
much more than neutrino experiments. Searches for 
\obb~are fundamental physics because they probe the presence of 
lepton number violation, which is on equal footing to  
baryon number violation, i.e.~proton decay. 
Once a (positive or negative) result from \obb~experiments is present,
one can go on and (assuming that lepton number is violated in Nature) 
 interpret the outcome in two ways:
\begin{enumerate}
\item {\it Standard Interpretation}: 

neutrinoless double beta decay is mediated by light, active and massive
Majorana neutrinos (the ones which oscillate) and all other mechanisms
potentially leading to \obb~give negligible or no contribution; 

\item {\it Non-Standard Interpretations}:

neutrinoless double beta decay is mediated by some other lepton number
violating process, and light, active and massive
Majorana neutrinos (the ones which oscillate) potentially leading 
to \obb~give negligible or no contribution. 

\end{enumerate}

In the first interpretation \obb~is a neutrino physics experiment, in
the second one a broader particle physics experiment with emphasis on
the particular lepton number violating physics under study. 
Of course, the observation of \obb~implies the Majorana nature
of neutrinos (to be precise, a tiny 4-loop induced Majorana mass term), 
a fact known as the black-box, or Schechter-Valle, theorem
\cite{Schechter:1980gr}. 
 
We will discuss in this contribution some of the  
physics potential of the two interpretations given above. For experimental
aspects, see the contributions in \cite{exp}, nuclear physics issues 
are dealt with in \cite{nme}.

\section{\label{sec:s}Standard Interpretation}
\begin{figure}[t]
\vspace{-49pt}
\includegraphics[width=5.2cm]{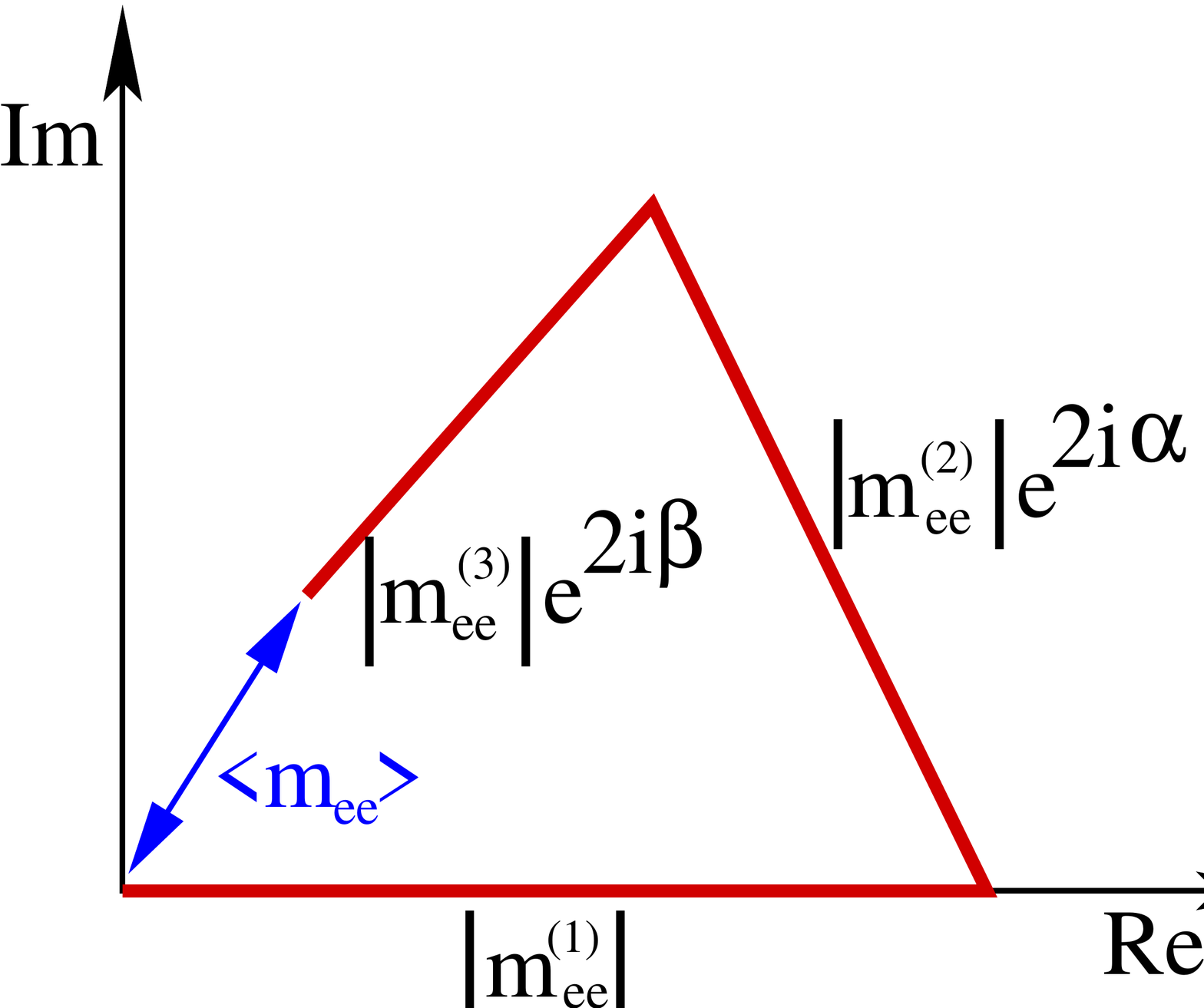}
\includegraphics[width=5.65cm]{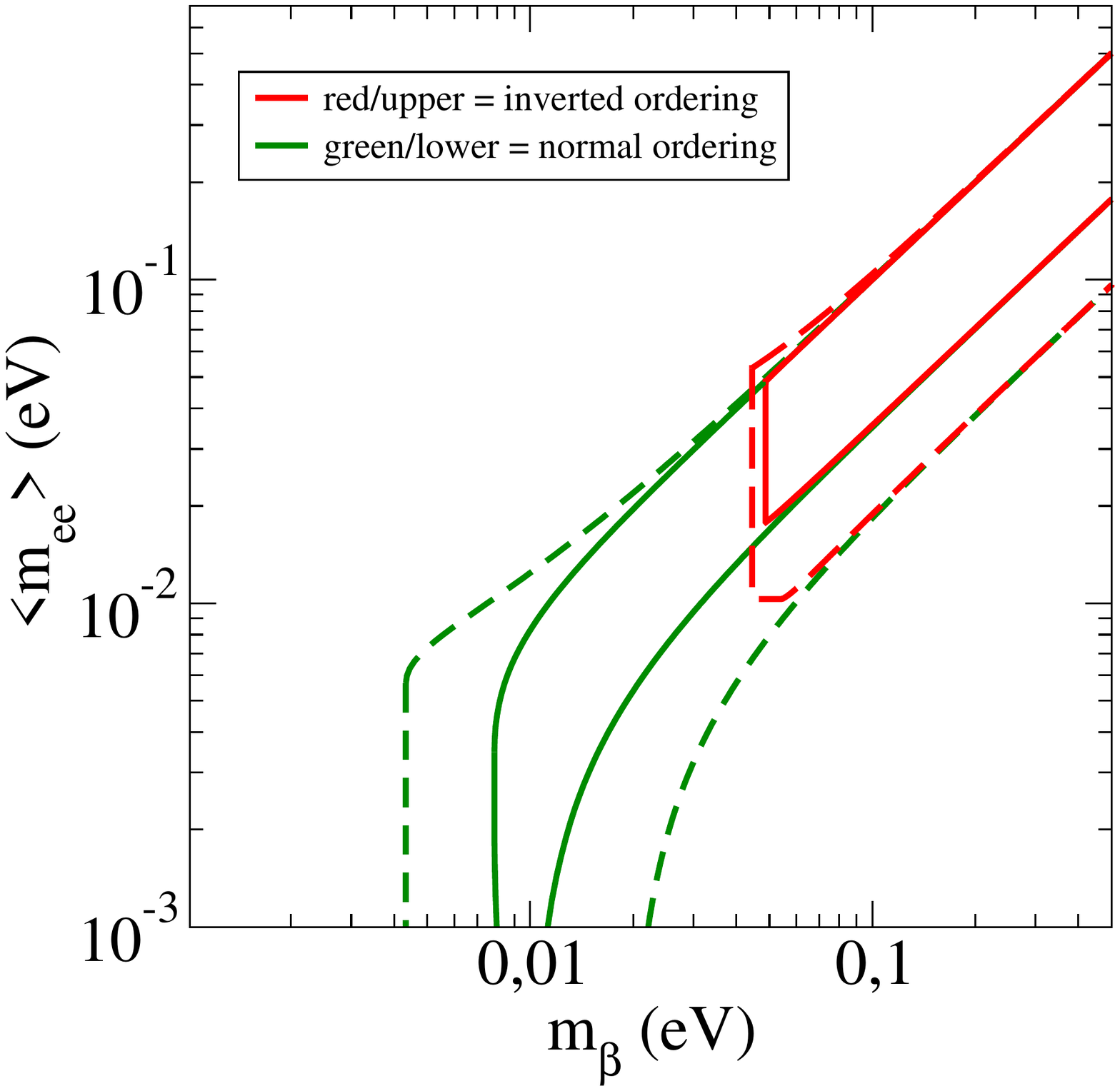}
\includegraphics[width=5.65cm]{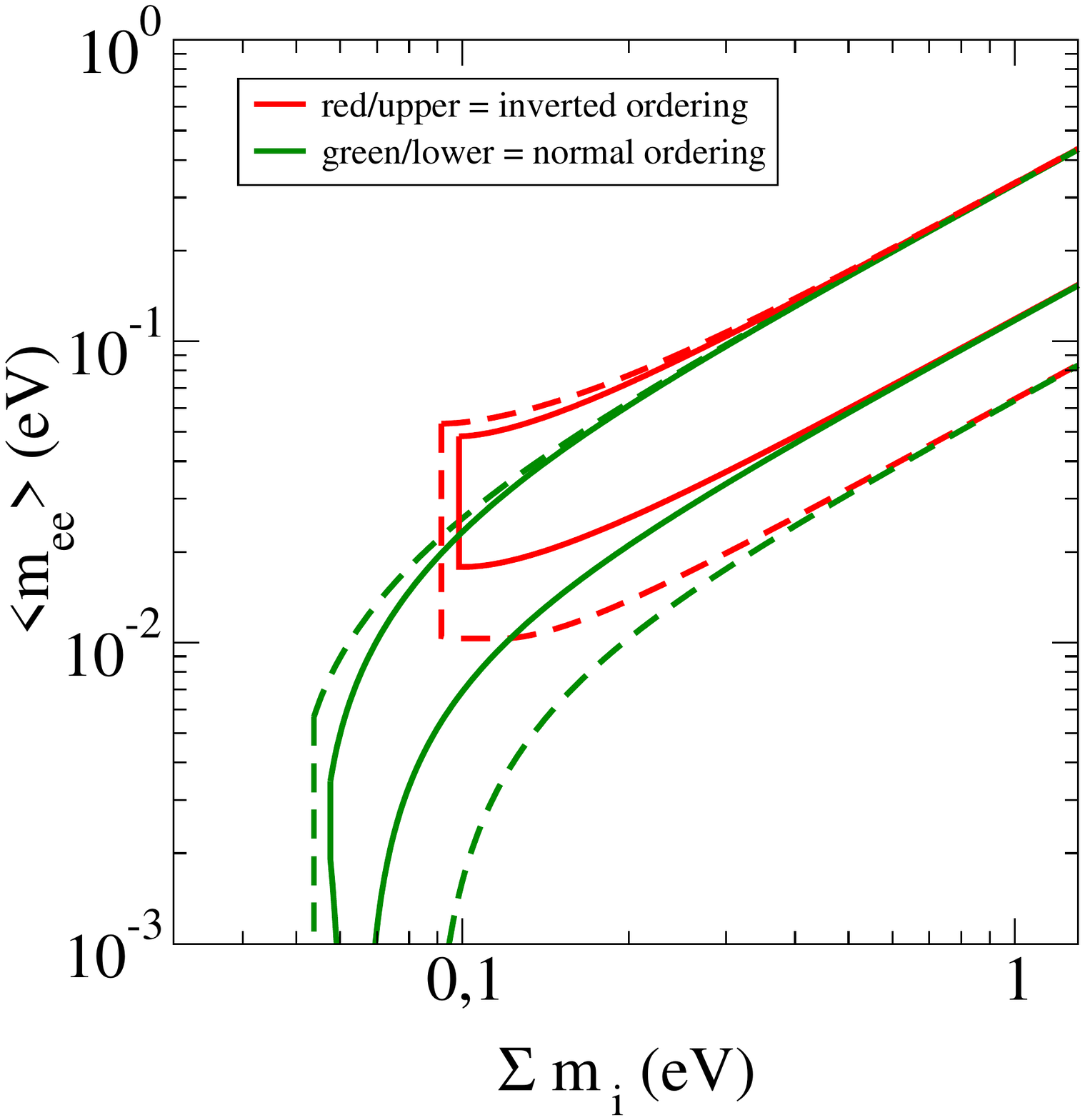}
\caption{Left: geometrical interpretation of the effective
mass. 
Middle: effective mass vs.~$m_\beta$. Right: effective mass
vs.~sum of masses. Both the best-fit values and the $3\sigma$ ranges
of the oscillation parameters are used.}
\label{fig:triangle}
\end{figure}

The first interpretation is most common, and in the light of
neutrino oscillations arguably the best motivated one. Assuming light massive
Majorana neutrino exchange, the amplitude for \obb~is proportional to 
\be \label{eq:ALNE}
{\cal A} \propto G_F^2 \, \frac{\meff}{q^2} ~\mbox{ with }~
\meff \equiv  \left| \sum  U_{ei}^2 \, m_i \right|
=  \left| |U_{e1}|^2 \, m_1 + |U_{e2}|^2 \, m_2 \, e^{2 i \alpha} 
+ |U_{e3}|^2 \, m_3 \, e^{2 i \beta} \right| \, . 
\ee 
Here $q^2 \simeq (0.1~{\rm GeV})^2$ is the typical momentum exchange in the
reaction, $m_i$ are the neutrino masses, $|U_{ei}|$ the PMNS matrix elements
of the first row (depending on $\theta_{12}$ and $\theta_{13}$) and 
$\alpha, \beta$ are the two Majorana phases. The coherent sum \meff~is
usually called the effective mass and contains 7 of 9 parameters of
the neutrino mass matrix, which in the fundamental
Lagrangian fully describes neutrino mass and lepton mixing. 
Two of those 9 parameters, the Majorana
phases, show up only in the effective mass. 
It contains therefore a large amount of
information. Its geometrical interpretation is shown in the left part
of Fig.~\ref{fig:triangle}: if the three complex terms in $\sum
U_{ei}^2 \, m_i$ cannot
form a triangle, the effective mass is non-zero. 
Fig.~\ref{fig:triangle} also displays plots of the effective mass
versus the other two complementary mass observables. Those are 
\be \label{eq:m_other}
m_\beta = \sqrt{\sum |U_{ei}|^2 \, m_i^2}~\mbox{ and }~
\sum m_i = m_1 + m_2 + m_3 \, ,  
\ee
measurable in beta decays \cite{katrin} and in cosmology
\cite{cosmo}, respectively. We refer to the cited contributions presented at this
conference for their current status and future prospects. 

What is noteworthy from the plots of neutrino mass observables is
that the effective mass can vanish in case of the normal neutrino mass
ordering. While it seems at first sight unnatural that 7 parameters
conspire in order to let a particular combination of them
become zero, it should be kept in mind that \meff~is the $ee$ element
of the fundamental low energy Majorana neutrino mass matrix. 
This matrix is 
generated by the underlying theory of mass generation, and texture
zeros occur frequently in such (flavor) models. We recall here that the
dependence on \meff~of the amplitude is in fact stemming from the 
term $\sum U_{ei}^2 \, m_i/(q^2 - m_i^2)$, and the limit $q^2 \gg
m_i^2$ is taken in the fermion propagator to obtain Eq.~(\ref{eq:ALNE}). 
If $\sum U_{ei}^2 \, m_i = 0$, then there is a term of order
$U_{ei}^2 \, m_i^3/q^4$, which will in general 
not vanish. It is however
suppressed by a factor $m_i^2 /q^2 \ls 10^{-16}$. 

In contrast to the normal ordering, the effective mass cannot vanish for the inverted mass
ordering. The lower limit of \meff~can be expressed as 
\be
\meff^{\rm IH}_{\rm min} = \left(1 - |U_{e3}|^2\right)  
\sqrt{|\dma|}  \left( 1 - 2 \sin^2 
\theta_{12} \right) .  
\ee
If the inverted ordering is to be ruled out, limits below $\meff^{\rm IH}_{\rm
min}$ have to be reached.
Among the parameters governing $\meff^{\rm IH}_{\rm min}$ the 
largest dependence is induced by the
solar neutrino parameter $\sin^2 \theta_{12}$, and quantifies to an
uncertainty of about a factor of 2 for $\meff^{\rm IH}_{\rm min}$. 
This factor of 2 introduces therefore 
about the same uncertainty as the nuclear matrix elements (NMEs), and 
motivates solar neutrino precision experiments in order to reduce it.

The three types of neutrino mass observables are obviously highly
complementary. 
Ordinary beta decay is basically a model-independent probe of neutrino mass, whereas the
extraction of the cosmological observable $\sum m_i$ is sensitive to
the data sets used, and little is known what happens to limits when
non-standard cosmological models different from the $\Lambda$CDM
framework are applied. As mentioned above, neutrino mass limits from
\obb~require the assumption of lepton number violation and in
addition, as we will see below, that no other mechanism contributes. 

\begin{center}
\begin{figure}[t]
\vspace{9pt}
\includegraphics[width=5.46cm,height=5cm]{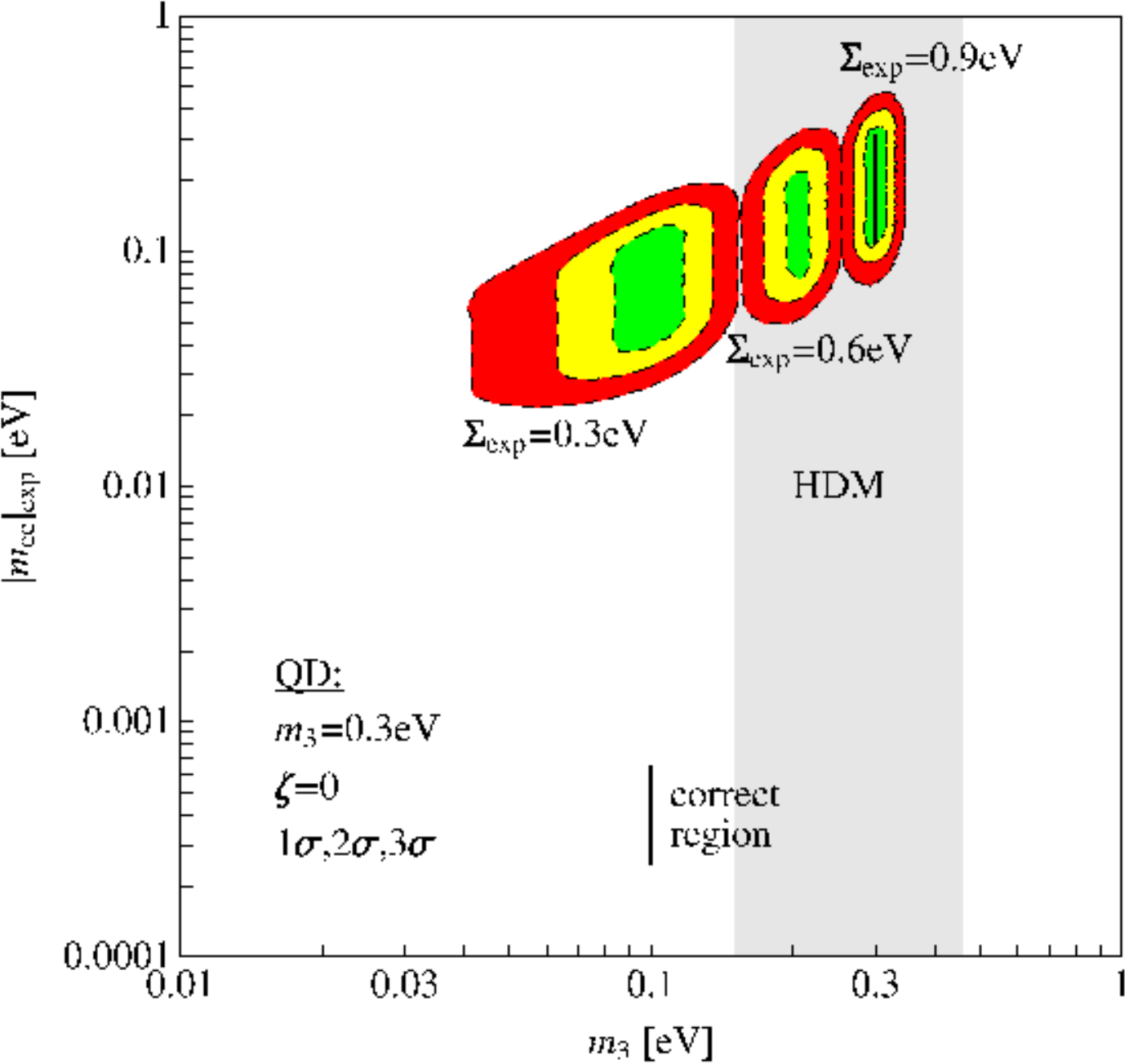}
\includegraphics[width=5.46cm,height=5cm]{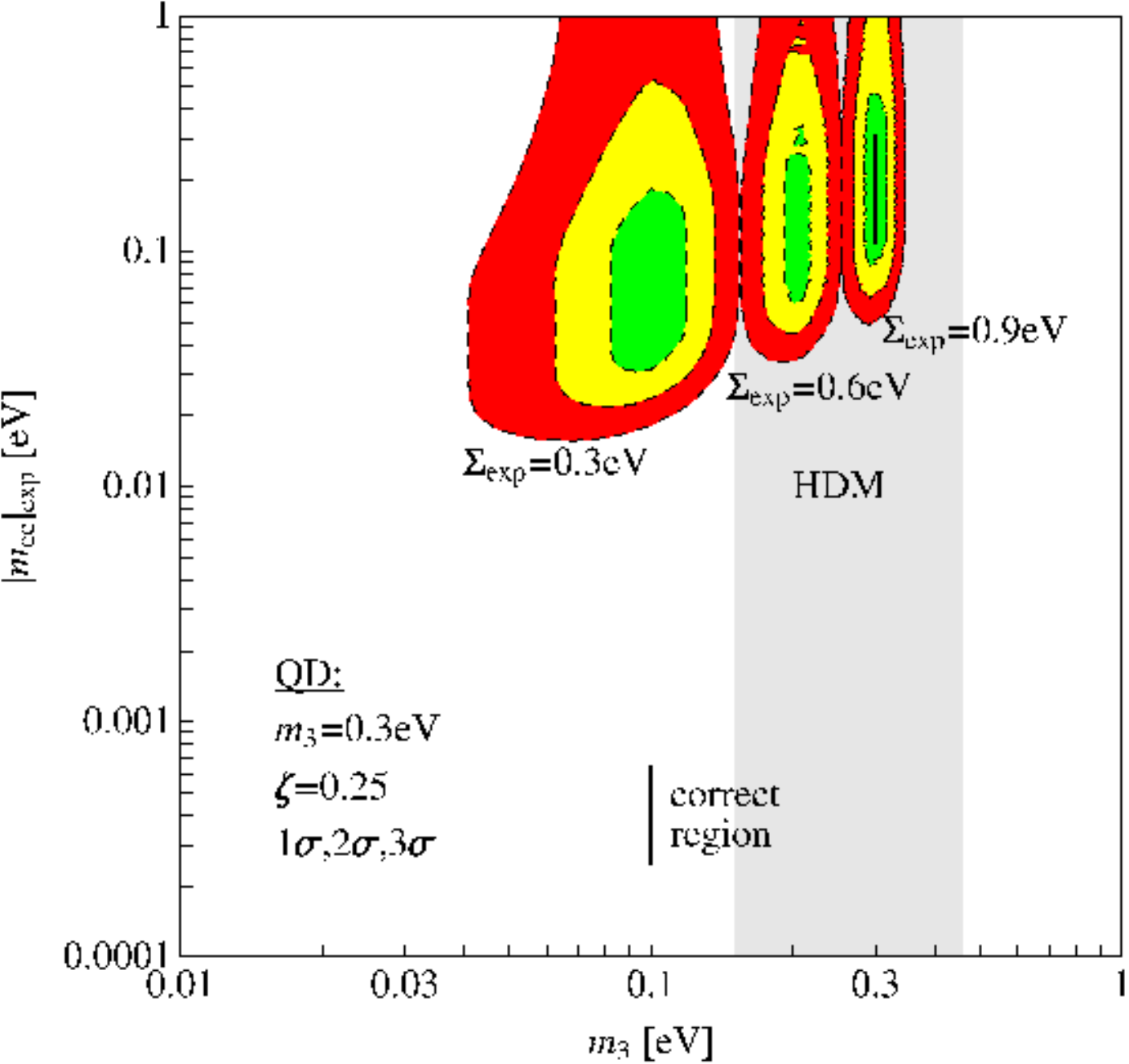}
\includegraphics[width=5.46cm,height=5cm]{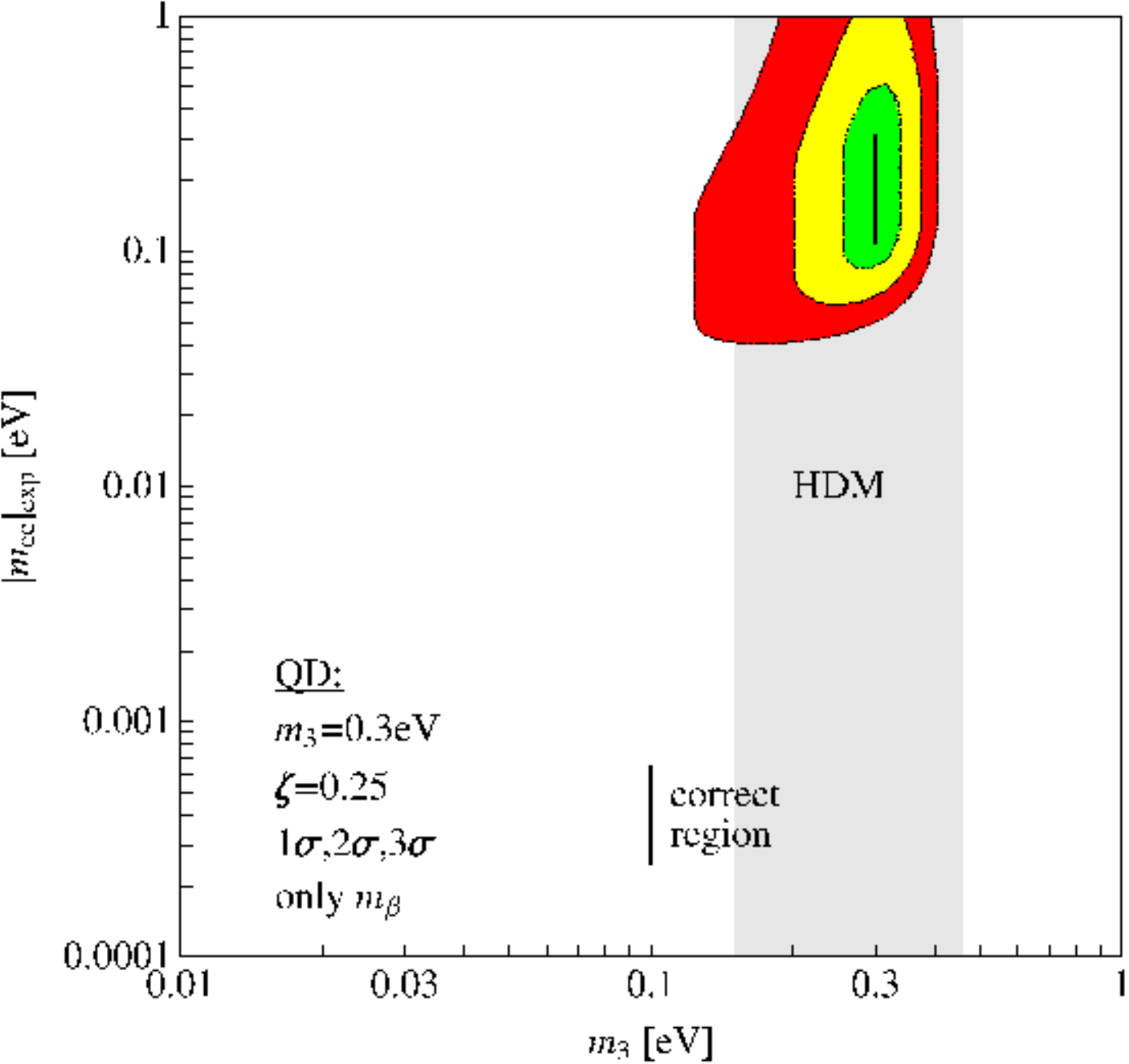}
\caption{1, 2 and 
3$\sigma$ regions in the $m_3$-$\meff_{\rm exp}$ plane for a true
value of $m_3 = 0.3$ eV. The solid line is the correct region. 
The left plot is for no NME uncertainty, the middle and right plots 
for $\zeta = 0.25$. Three different measured values of $\Sigma$
are assumed for the left and middle plots. In the right plot the
cosmological mass limit is left out of the analysis. 
The area denoted HDM is the range of \meff~from the claim of 
part of the Heidelberg-Moscow collaboration. Taken from \cite{Maneschg:2008sf}.}
\label{fig:stat}
\end{figure}
\end{center}

\begin{center}
\begin{figure}[t]
\vspace{9pt}
\hspace{.1cm}\includegraphics[width=7.41cm,height=7cm]{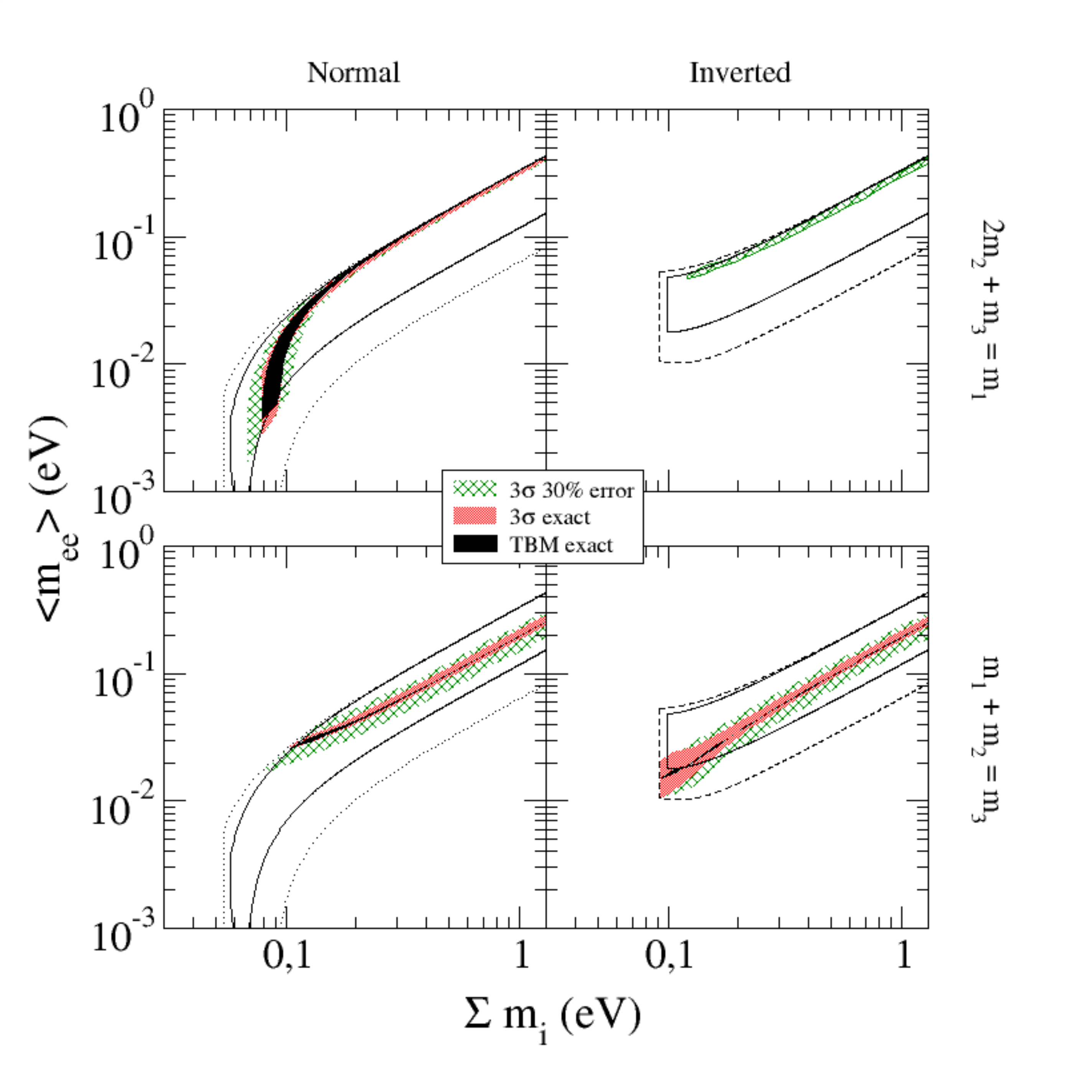}\hspace{1cm}
\includegraphics[width=7.41cm,height=7cm]{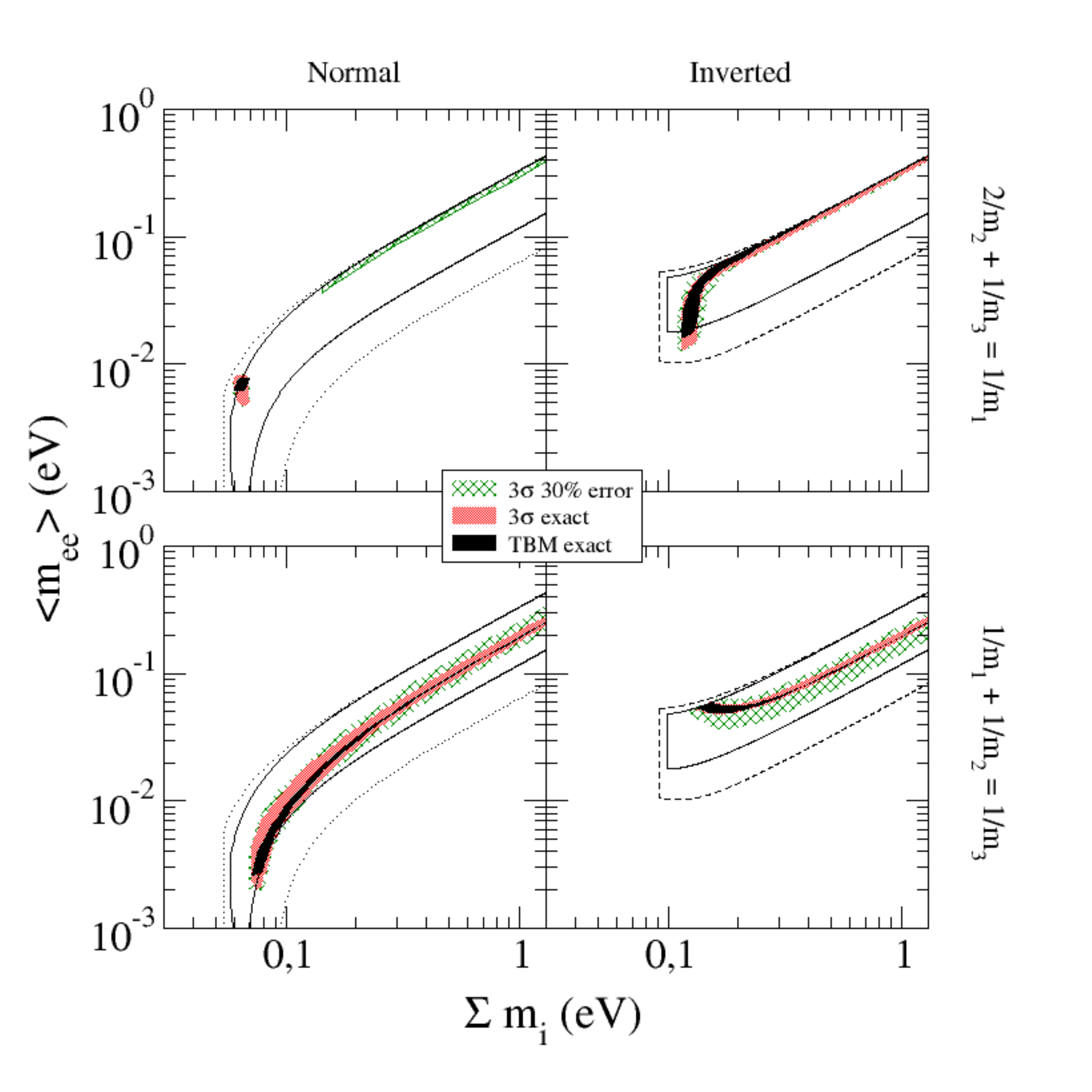}
\caption{Allowed regions in $\meff - \sum m_i$ parameter space for the
sum-rules $2m_2+m_3=m_1$ (top left), $m_1+m_2=m_3$ (bottom left), 
$\frac{2}{m_2}+\frac{1}{m_3}=\frac{1}{m_1}$ (top right) and 
$\frac{1}{m_1}+\frac{1}{m_2}=\frac{1}{m_3}$ (bottom right). The
regions allowed in the general case are indicated by the black lines. 
Neutrino masses are here understood to be complex. Taken from \cite{Barry:2010yk}.}
\label{fig:sr}
\end{figure}
\end{center}

At the end of this decade, $m_\beta$ will be 
known to be larger or smaller
than about 0.3 eV, standard cosmology will be sensitive 
to smaller values of $\sum m_i$, while \meff~can also be probed down to the 0.1 
eV regime. All three observables are therefore expected to be
tested at similar levels, and the complementarity of the different 
approaches to neutrino mass opens up exciting possibilities. 
The interplay of the observables is studied in detail e.g.~in 
Refs.~\cite{Feruglio:2002af,deGouvea:2005hj,Lindner:2005kr,Pascoli:2007qh,Fogli:2008ig}.
The ideal case would arise when positive signals in all three
measurements were found. Using the foreseen experimental uncertainties
(and theoretical uncertainties, i.e.~the NMEs) one
can estimate how precisely the neutrino mass scale could be pinned down 
\cite{Pascoli:2005zb,Maneschg:2008sf}. 
The following analysis, taken from Ref.~\cite{Maneschg:2008sf} (similar
results can be found in \cite{Pascoli:2005zb}),
assumes quasi-degenerate neutrinos with a true value of the neutrino
mass of $m_3 = 0.3$ eV (hence $m_\beta = 0.3$ eV and $\sum m_i = 0.91$
eV) and experimental errors as specified in publications of the 
respective collaborations. 
The remaining free parameter is called $\zeta \ge 0$ and quantifies
the uncertainty introduced by our ignorance about the NMEs in the
extraction of \meff~from a lifetime measurement. Without any NME
uncertainty we could determine $m_3$ at $3\sigma$ to
about 15 \%, while $\zeta = 0.25$ would make this possible to 25
\%. Leaving $\sum m_i$ out of the analysis would lead to an error of 50 \% on
$m_3$ if $\zeta = 0.25$, which illustrates that the precision is
largely from the determination of $\sum m_i$. 
 This is all illustrated in Fig.~\ref{fig:stat},
which displays the reconstructible 1, 2 and 3$\sigma$ contours in the
parameter space of $m_3$ and the measured effective mass $\meff_{\rm
exp}$. \\
Another aspect of mass-related observables is the potential to rule
out some of the many models which have been proposed to explain lepton
mixing. Very often flavor symmetry models generate neutrino mass
sum-rules (see e.g.~\cite{mod})
and thereby generate relations between the observables that are different from
the general case (Fig.~\ref{fig:triangle}) and therefore only
certain regions in parameter space are allowed. Fig.~\ref{fig:sr},
taken from Ref.~\cite{Barry:2010yk}, shows the result for four sum-rules.

Before turning to non-standard interpretations, i.e.~mechanism of 
double beta decay not directly connected to 
3 neutrino oscillations, let us mention an ``intermediate case'': if light 
sterile neutrinos exist, which could be suggested by interpretations of the 
LSND/MiniBooNE experiments (see \cite{st} for their current status), then there are 
eV scale sterile neutrinos with mixing angles of order 0.1. 
Hence, their contribution to the effective mass \cite{sruba} is at
least of order 0.01 eV, i.e.~of the same order as for the inverted
hierarchy.


\section{\label{sec:ns}Non-Standard Interpretations}
A clear experimental signature for a non-standard contribution to
\obb~would be for instance no signal in KATRIN and/or cosmology, and a
life-time measurement in \obb~which would be interpreted as an
effective mass of 0.5 eV or so. There are several candidates for
non-standard contributions: Higgs triplets, right-handed currents,
heavy Majorana neutrinos, supersymmetric particles, etc. Limits from
\obb~can be translated into limits on the couplings and masses
associated with these mechanisms. The limits from the
literature, which often take nuclear physics aspects into account, 
can be approximately reproduced by rather simple arguments 
on the amplitude level. 
For instance, heavy Majorana neutrinos (for which in the propagator 
the limit $M_i^2 \gg q^2$ applies) will have 
an amplitude proportional to ${\cal A} \propto G_F^2 \, \frac{S_{ei}^2}{M_i}$,  
where $M_i$ are heavy neutrino masses with coupling $S_{ei}$ to
electrons. Setting this amplitude equal to the light neutrino
amplitude from Eq.~(\ref{eq:ALNE}), and using $\meff \ls 0.5$ eV gives
$S_{ei}^2/M_i \ls 5 \times 10^{-8}$ GeV$^{-1}$, which is in fact the
limit given in \cite{KlapdorKleingrothaus:1999kq}. (Left-handed) Higgs triplet
exchange has an amplitude ${\cal A} \propto G_F^2 \, h_{ee} \, v_L /M_\Delta^2$, where
$h_{ee}$ is its coupling to two electrons, $v_L$ its vev, and
$M_\Delta \gg q$ its mass. Note that $h_{ee} v_L$ is the
contribution of the triplet to neutrino mass. Therefore, 
if the triplet would be responsible for neutrino mass, its contribution 
to \obb~would be suppressed by a factor 
$q^2/M_\Delta^2$. On the other hand, the triplet can only give the leading 
contribution to \obb~if \meff~is extremely small. 

$R$-parity violating SUSY can
also mediate the process (see e.g.~\cite{Mohapatra:1986su,Hirsch:1995ek}), 
in two classes of diagrams. First, in the
usual diagram of \obb~the $W$ bosons can essentially 
be replaced by selectrons and the
Majorana neutrino by a neutralino or gaugino. The amplitude is then given by 
${\cal A} \propto g^2 \, \frac{{\lambda'}_{111}^2}{\Lambda_{\rm SUSY}^5}$, where $g
\simeq \sqrt{0.1}$ is a (combination of) gauge coupling(s), 
$\lambda'_{111}$ stems from the vertex of the selectron with the up-
and down quarks and the power of $\Lambda_{\rm SUSY}$ is easily
understood by the propagators for one fermion and two bosons, whose masses are
assumed to be $\Lambda_{\rm SUSY} \gg q$. 
By comparing again with the amplitude in 
Eq.~(\ref{eq:ALNE}), it follows 
$\frac{{\lambda'}_{111}^2}{\Lambda_{\rm SUSY}^5} \ls 7 \times 10^{-17}$
GeV$^{-5}$, to be compared with the literature value of $3 \times 10^{-17}$
GeV$^{-5}$ \cite{Hirsch:1995ek}. Interestingly, the same couplings 
describe resonant selectron production at the LHC \cite{Allanach:2009iv}, 
which allows to test this mechanism\footnote{LHC related
phenomenology of non-standard mechanisms in 
left-right symmetric theories has recently been discussed in
\cite{Tello:2010am}.}. 
For instance, one can show that 
certain regions in SUSY parameter space are in conflict
with existing \obb-limits, or that detection of resonant selectron 
production at the LHC in other regions would
rule out any considerable contribution of this mechanism 
to \obb~\cite{Allanach:2009iv}. 
Another class of $R$-parity violating diagrams involves neutrino and virtual
squark exchange. The amplitude goes as 
${\cal A} \propto G_F \, m_{d_k} {\frac{{\lambda'}_{1k1} \,{\lambda'}_{11k}}
{q \, \Lambda_{\rm SUSY}^{3}}}$, where $m_{d_k}$ is the down-type quark mass of
the $k$th generation. It enters the game because mixing between
left- and right-handed squarks is involved, which is proportional to
this mass. The ${\lambda'}_{121} \,{\lambda'}_{112}$ term is 
irrelevant due to $K^0$-$\bar{K}^0$ constraints, and the
${\lambda'}_{111} \,{\lambda'}_{111}$ contribution is sub-leading with
respect to the diagram discussed above \cite{Allanach:2009xx}. 
Anyway, depending on whether it is the down-, strange- or bottom
(s)quark, by comparing the amplitudes 
limits of ($1 \times 10^{-11}$, $6 \times 10^{-13}$ or $1 \times
10^{-14}$) GeV$^{-3}$ arise, compared with the actual literature values
of ($7.7 \times 10^{-12}$, $4.0 \times 10^{-13}$ or $1.7 \times
10^{-14}$) GeV$^{-3}$ \cite{Faessler:2007nz}. Again, the simple
estimates are rather close to the limits involving nuclear physics. 

However, there is nuclear physics, and in fact it can help to
distinguish the different mechanisms. This has been dealt with
recently in
Refs.~\cite{Gehman:2007qg,Deppisch:2006hb,Fogli:2009py}. For instance,
one can within a typical NME calculation fix the particle physics
parameters such that for $^{76}$Ge the life-time is the same for all
popular non-standard mechanisms. Triggered by nuclear details, the
life-time in other nuclei can however differ by up to one order of
magnitude. Typically, multi-isotope determination of \obb~in three to
four different elements is necessary in order to single out the true
mechanism \cite{Gehman:2007qg}. 

\begin{center}
\begin{figure}[t]
\vspace{9pt}
\hspace{1cm}\includegraphics[width=6.cm,height=5cm]{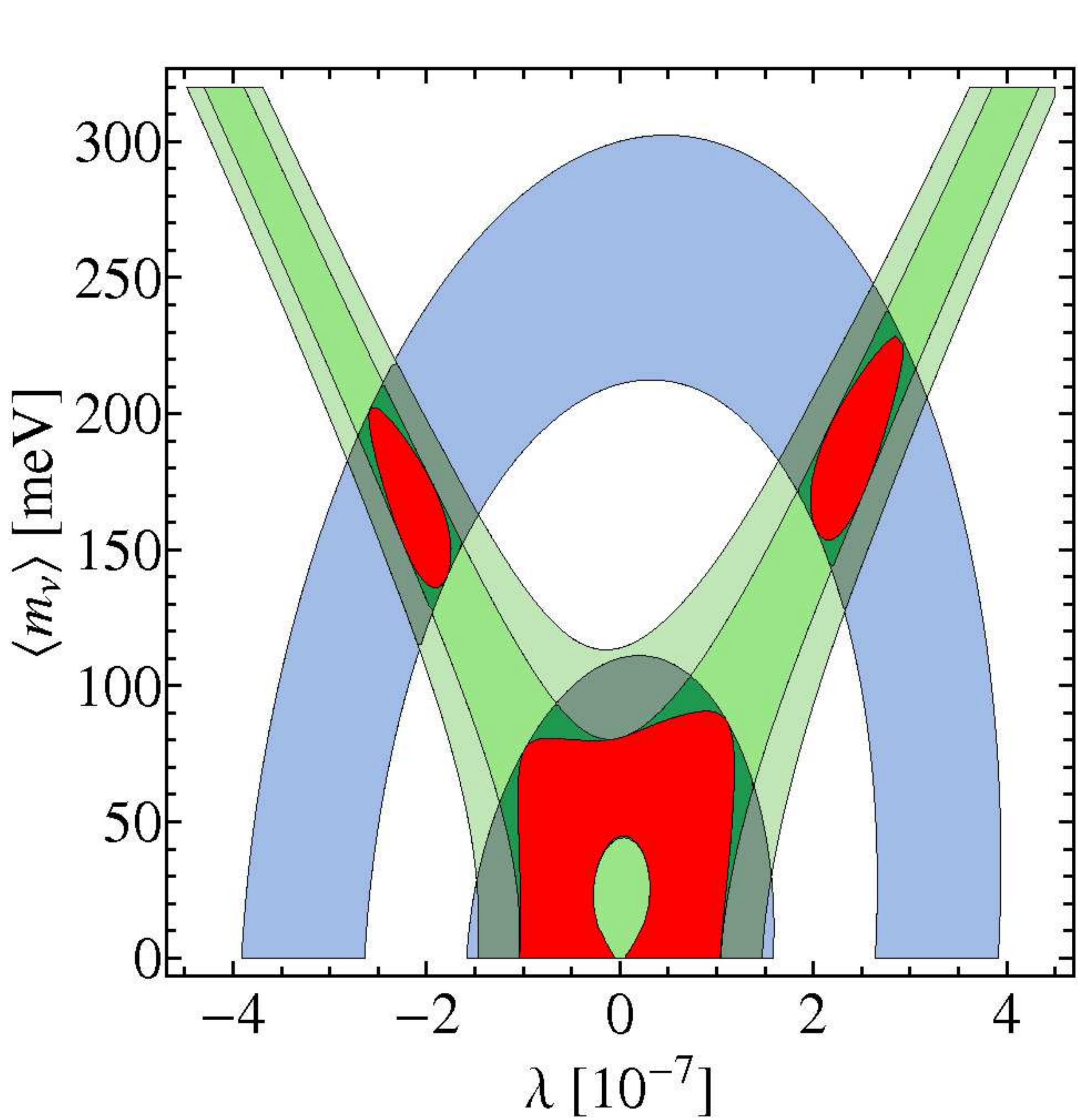}\hspace{1cm}
\includegraphics[width=6.cm,height=5cm]{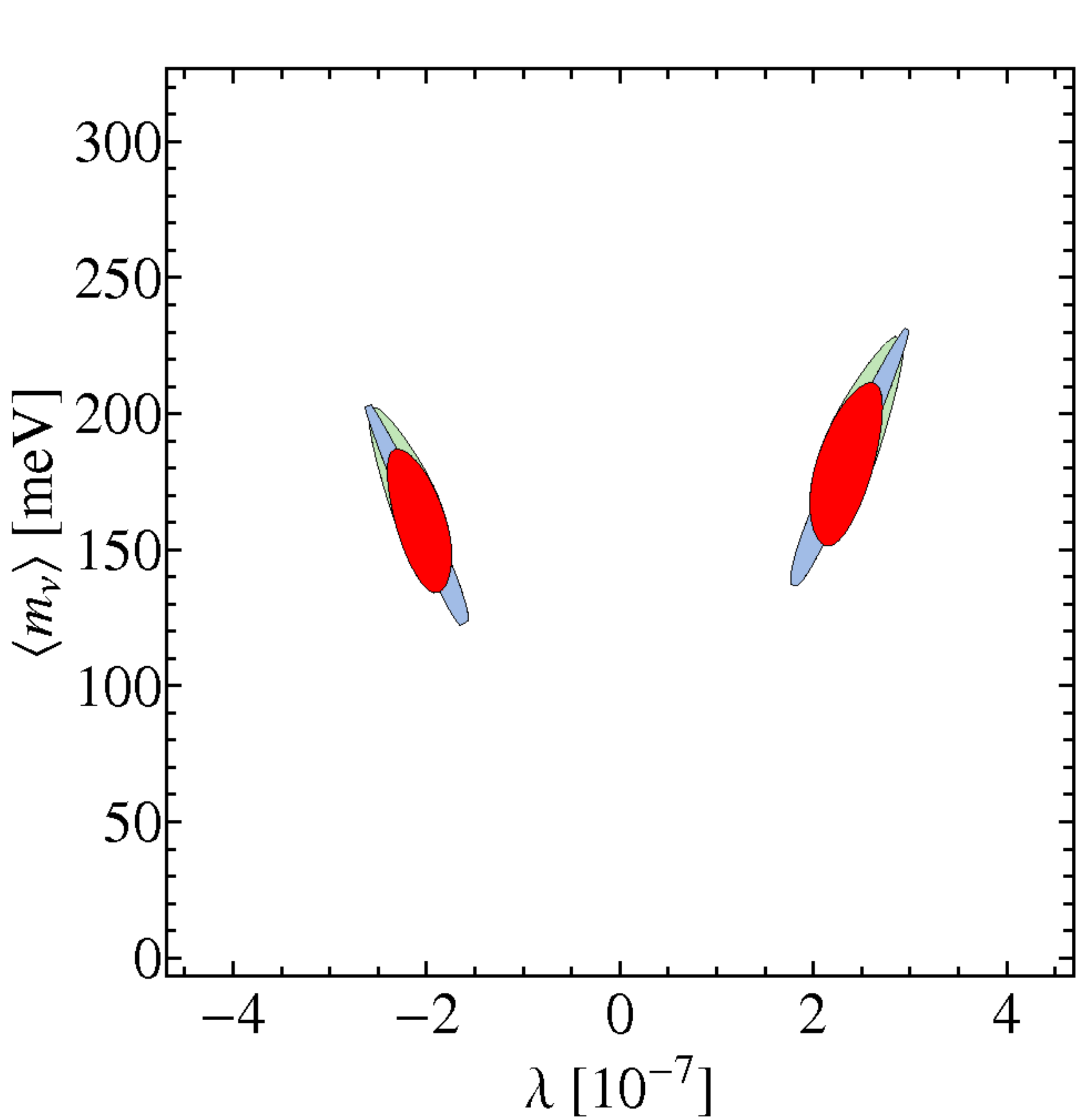}
\caption{Left: constraints at $1\sigma$ on the model parameters from 
an observation of \obb~of $^{82}$Se at half-life $10^{25}$ y 
(outer blue elliptical area) and $10^{26}$ y (inner blue elliptical
area). Adding the reconstruction of the angular (outer, lighter green)
and energy difference (inner, darker green) distribution drastically
shrinks the allowed parameter space. Right: adding information from
the decay of $^{150}$Nd. In this example, 30 \% admixture of
right-handed currents is assumed. Taken from \cite{Arnold:2010tu}.}
\label{fig:sn}
\end{figure}
\end{center}

We have seen up to know that non-standard mechanisms can be pinned down
either by looking for other places in which they show their presence,
or by probing \obb~in different nuclei. The third possibility is to
take a closer look at the decay products, namely the two
final state electrons. The SuperNEMO experiment is currently the only
one able to probe in particular the energy of the individual electrons
and their angular distribution \cite{Arnold:2010tu}. A recent analysis
by the collaboration assumes the simultaneous presence of the standard
mechanism (with the particle physics parameter \meff) and a 
right-handed current contribution 
($\lambda = (m_W/m_{W_R})^2 \, U_{ei} \, V_{ei}$, with $V$ being the right-handed 
analogon of the PMNS matrix). Fig.~\ref{fig:sn}
shows an example of the results from \cite{Arnold:2010tu}.

\section{Summary}
Neutrinoless double beta decay will be intensively searched for in
the current decade. The interesting complementarity with the other
neutrino mass-related observables can shed some light on the question
of neutrino mass generation and on the cosmological model. Perhaps
even more interesting is the possibility of other mechanisms leading to
\obb, which have phenomenological consequences in a variety of fields,
such as lepton flavor violation or accelerator physics.

\begin{center}
{\bf Acknowledgments}
\end{center}
This work was supported by the ERC under the Starting Grant 
MANITOP and by the DFG in the project RO 2516/4-1 as well as in the 
Transregio 27.

\bibliographystyle{elsarticle-num}
\bibliography{Proceedings_Rodejohann_Werner_Wednesday_I}


\end{document}